\documentclass[a4paper,11pt]{article}

\usepackage{pos}
\usepackage{xcolor}
\usepackage{siunitx}
\usepackage{amsmath}

\title{Galactic and Extragalactic Analysis of the Astrophysical Muon Neutrino Flux with 12.3 years of IceCube Track Data}
\ShortTitle{IceCube: Galactic and Extragalactic Analysis of Astrophysical Muon Neutrino Flux}

\author{The IceCube Collaboration \\{\normalsize \normalfont(a complete list of authors can be found at the end of the proceedings)}\\}

\emailAdd{philipp.fuerst@icecube.wisc.edu}

\abstract{
The Ice Cube Neutrino Observatory has been measuring an isotropic astrophysical neutrino flux in multiple detection channels for almost a decade. Galactic diffuse emission, which arises from the interactions between cosmic rays and the interstellar medium, is an expected signal in IceCube. The superposition of an extragalactic flux and a galactic flux results in directional structure and variations in the spectrum. In this work, we use 12.3 years of high-purity muon-neutrino induced muon track data to perform a dedicated search for this galactic emission, combined with a spectral measurement of the isotropic astrophysical neutrino flux. To distinguish a galactic component from the dominant atmospheric and isotropic astrophysical components, the precise directional information available for muon tracks is fully utilized in a three-dimensional forward folding likelihood fit. We test a state-of-the-art model prediction of galactic diffuse emission based on recent cosmic ray data (CRINGE). We fit this prediction as a template scaled by a factor $\Psi_{\mathrm{CRINGE}}$, and find $2.9\pm 1.1 \times \Psi_{\mathrm{CRINGE}}$ with a significance of $2.7\sigma$ in an energy range between \SI{400}{GeV} and \SI{60}{TeV} in the Northern Sky. 

\vspace{4mm}
{\bfseries Corresponding authors:}
Philipp Fürst$^{1}$\\
{$^{1}$\itshape RWTH Aachen University}\\
\ConferenceLogo{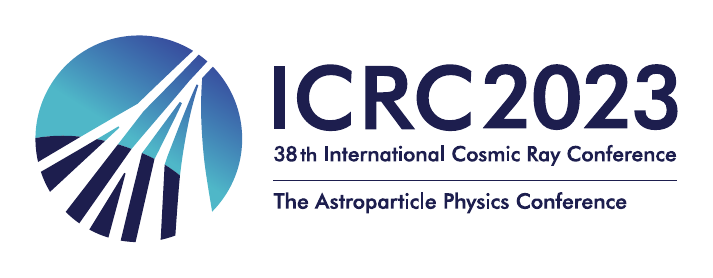}
\FullConference{The 38th International Cosmic Ray Conference (ICRC2023)\\ 26 July -- 3 August, 2023\\ Nagoya, Japan}
}

\begin{document}
\maketitle
\section{Introduction}\label{sec:intro}
Galactic diffuse neutrino emission from hadronic interactions between cosmic rays and the interstellar medium of our galaxy can serve as a direct tracer for cosmic rays and is an expected signal for IceCube. Neutrino emission from the galactic plane has very recently been observed for the first time with high significance by IceCube \citep{dnncascades, mircosteve:2023icrc}, emerging as a non-isotropic, subdominant component from the long-observed diffuse astrophysical neutrino flux.

Here, we present an update to one of the analyses investigating this isotropic diffuse flux \cite{10diffuse}. We use 12.3 years of track data from the northern hemisphere and extend the analysis variables of reconstructed energy proxy and zenith angle by a third dimension by including right ascension information. This enables a combined measurement of both the isotropic flux, identified by its high-energy signature where it dominates over atmospheric backgrounds, and a galactic signal, identified by its unique spatial structure on the sky. Atmospheric backgrounds are completely isotropized in right ascension due to the rotation of the Earth. The baseline signal hypothesis is formulated as an isotropic single powerlaw and an additional component from galactic diffuse emission \cite{cringe}, see Sec. \ref{sec:analysis}. Additionally, different model calculations for galactic diffuse emission are tested \citep{fermipi0, KRAgamma, FangMuraseModels}, providing hints about the structure of this galactic emission.

\section{IceCube Detector and Selection of Data Sample}\label{sec:data}
The IceCube Neutrino Observatory was built to detect astrophysical neutrinos, which it discovered in 2013 \cite{icecube_firstastro}. It has been measuring their spectrum in multiple detection channels since, with the most recent results presented at this conference \citep{globalfit:2023icrc, estes:2023icrc, mese:2023icrc}.

Charged secondaries produced by deep inelastic scattering (\textit{DIS}) of neutrinos and matter produce a detectable signature of Cherenkov light. IceCube instruments about \SI{1}{km}$^3$ of natural ice at the South Pole to detect this light, embedding 5160 digital optical modules (DOMs) in depths between \SI{1.5}{km} and \SI{2.5}{km} in the ice. The DOMs are mounted on a total of 86 vertical cables (strings) \cite{icecube_jinst}. Possible secondaries of \textit{DIS} include high-energy muons, which can travel up to several kilometers in ice, creating a track-like signature in the detector. 

This analysis is based on a selection of such track-like events. It utilizes boosted decision trees trained to distinguish high-quality tracks from spherical cascade topologies (which occur for example from $\nu_e$ interactions) and a background of tracks induced by atmospheric muons, which dominate the flux from the Southern Sky. A zenith cut of $\Theta_{\mathrm{zenith}} > 85^\circ$ is applied, so only events which travelled through the Earth or substantial amounts of Antarctic ice and rock are accepted. This effectively shields the sample from muons created in the atmosphere by cosmic ray interactions, resulting in a neutrino purity > 99.8\%, at the cost of removing the hemisphere including the galactic center. The analysis region includes the section of the galactic plane with longitudes $<-141^{\circ}$ and $>27^{\circ}$.

This event selection has been used for measurements of the astrophysical neutrino flux \citep{tracks_6yr,10diffuse}, including a test for a galactic plane contribution \cite{chaackthesis}. Here, we extend the data sample to a total of 12.3 years of data collected between June, 2010 and January 2023, resulting in a 45\% increase in event numbers compared to the previous analysis \cite{10diffuse}. Data-taking seasons when the detector was running in partially-completed configurations such as IC59 (with 59 strings deployed) are excluded in favor of a unified detector simulation. The IC79 season has been shown to be sufficiently similar to the full IC86 configuration when weighted with a detection efficiency of 94\% \cite{10diffuse} and is included. This yields a total of 982,279 up-going, track-like events with a median angular resolution of $<1^\circ$ above \SI{1}{TeV} and $<0.25^\circ$ above \SI{1}{PeV}.

\section{Analysis Method}\label{sec:analysis}
The analysis is based on a forward-folding, three dimensional binned likelihood fit. The analysis dimensions are a muon energy proxy, \textit{truncated energy} \cite{abbasiImprovedMethodMeasuring2013}, and two angles of reconstructed event direction. The direction is calculated using the \textit{MPE} algorithm \citep{ahrensMuonTrackReconstruction2004} in coordinates of IceCube zenith $\Theta$ and right ascension RA. Data events are sorted into their respective bins (50 in energy proxy between \SI{100}{GeV} and \SI{10}{PeV}, 33 in $\cos{\Theta}$ and 180 in RA), yielding $n_{\mathrm{bin}}$ data events per bin, and are compared to the expectation $\mu_{\mathrm{bin}}$ calculated from simulated events. The expectation is a function of signal $\Vec{\Theta}$ and nuisance parameters $\Vec{\xi}$, and the maximum of the Poisson-likelihood $\mathcal{L}$ given the data $D$ yields the best-fit signal parameters:

\label{eq:llh}
\begin{equation}
    \mathcal{L}(D|\Vec{\Theta}, \Vec{\xi}) = \sum_{\mathrm{bin}}^{N_\mathrm{bins}} p_{\mathrm{Poisson}}(n_{\mathrm{bin}}, \mu_{\mathrm{bin}}(\Vec{\Theta}, \Vec{\xi})).
\end{equation}

The nuisance parameters $\Vec{\xi}$ consist of two groups, the first one describes overall detector uncertainties and the second the background fluxes. The parameterization closely follows a previous analysis of this data sample using 9.5 years of data \cite{10diffuse}. 

Detector uncertainties arise from the overall light collection efficiency of the DOMs and absorption and scattering properties of the glacial ice. Additionally, the effects of impurities and bubbles produced by the re-freezing of water in the boreholes are included. These are modeled as an acceptance function depending on incident angle for the DOMs. Compared to \cite{10diffuse}, a second parameter describing this angular acceptance function has been added, affecting the possible zenith distribution of the fitted fluxes. This results in a total of five detector uncertainty parameters.

Across the sky, and for energies below $\approx$\SI{200}{TeV}, the dominating background arises from atmospheric neutrinos. Conventional and prompt atmospheric fluxes are updated compared to \cite{10diffuse} using the \textit{MCEq}-package \cite{mceq}, version 1.2.1. The cosmic ray primary flux is modeled with the \textit{H4a} model \cite{gaisserfluxH4a}, and for interactions in the atmosphere the hadronic interaction model \textit{Sibyll2.3c} is employed \cite{sibyll23c}. Uncertainty of the cosmic ray primary flux is covered by a combination of a free global spectral index shift and a parameter interpolating linearly between the \textit{H4a} and \textit{GST} \cite{gaisserfluxGST} models. The uncertainties of pion and kaon production in cosmic ray air showers are modeled following the \textit{Barr} formulation \cite{Barr}. A sub-dominant component from atmospheric muons which are mis-reconstructed as up-going is simulated using the \textit{CORSIKA} package \cite{corsika1998} and has a free normalization scale in the fit. In total, this formulation yields a number of $n_{\Vec{\xi}}=14$ nuisance parameters.

Two signal flux components are considered: an isotropic, astrophysical signal and a non-isotropic signal from the galactic plane. The isotropic signal is modeled as a single powerlaw (Eq. \ref{eq:spl}), and it is the only astrophysical component included in the baseline hypothesis $H_0$ (no galactic contribution). The galactic contribution is modeled following the CRINGE prediction for diffuse emission from \cite{cringe}. It is based on a fit to cosmic-ray data from multiple experiments and explores uncertainties arising from other required ingredients such as gas maps and cross sections. We preserve the complex spectral and spatial features of the prediction, which vary across the sky, but allow an overall free normalization scale $\Psi_{\mathrm{CRINGE}}$. This yields $n_{\Vec{\Theta}}=3$ signal parameters.

\section{Results}

\begin{figure}
    \centering
    \includegraphics[width=1.\columnwidth]{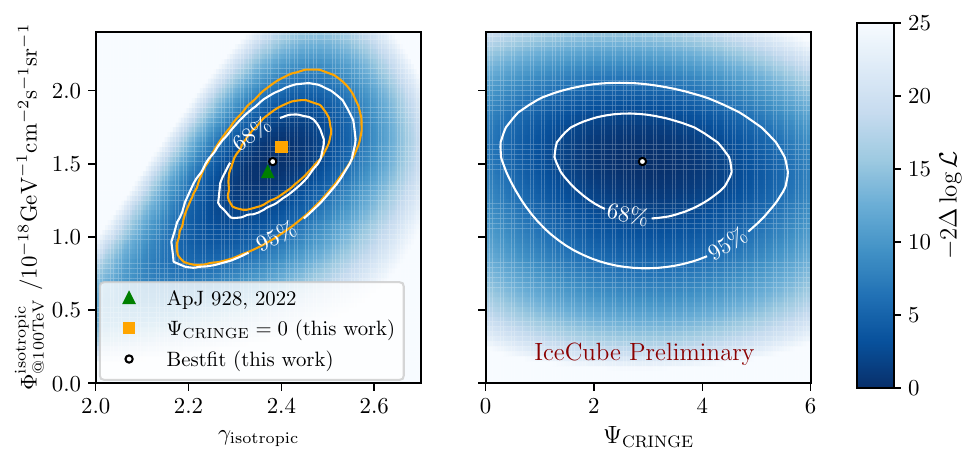}
    \caption{Likelihood landscape as a function of the signal parameters $\Phi_{\mathrm{isotropic}}$ ($y$-axis for both figures), $\gamma_{\mathrm{isotropic}}$, and $\Psi_{\mathrm{CRINGE}}$. The green triangle marks the result from \cite{10diffuse}, which did not consider a galactic contribution. The updated background calculations and removal of the IC59 data account for the observed shift towards the orange square, which is the 12.3 year fit without a galactic plane contribution. The best-fit normalization including a galactic component (white point) is 6.5\% lower.}
    \label{fig:contours}
\end{figure}

The null-hypothesis $H_0$ of no galactic contribution is excluded with a significance of $2.7\sigma$, and we fit a galactic contribution of $\Psi_{\mathrm{CRINGE}} =2.9\pm 1.1$ for our baseline model, which was chosen a priori. This template normalization corresponds to $1100\pm^{420}_{410}$ events, or 1.1\% of the total events. We fit an isotropic astrophysical component of $\phi_{\mathrm{isotropic}} = 1.51 \pm^{0.22}_{0.23}$ and $\gamma = 2.38\pm^{0.08}_{0.08}$ when described as a single powerlaw:

\begin{equation}
   \Phi^{\nu_\mu + \Vec{\nu}_\mu}_{\mathrm{isotropic}} =  \phi_{\mathrm{isotropic}} \times \left(\frac{E}{\SI{100}{TeV}}\right)^{-\gamma} \times 10^{-18}  \SI{}{GeV}^{-1} \SI{}{cm}^{-2} \SI{}{s}^{-1} \SI{}{sr}^{-1}. 
   \label{eq:spl}
\end{equation}

\begin{figure}
    \centering
    \includegraphics[width=0.78\textwidth]{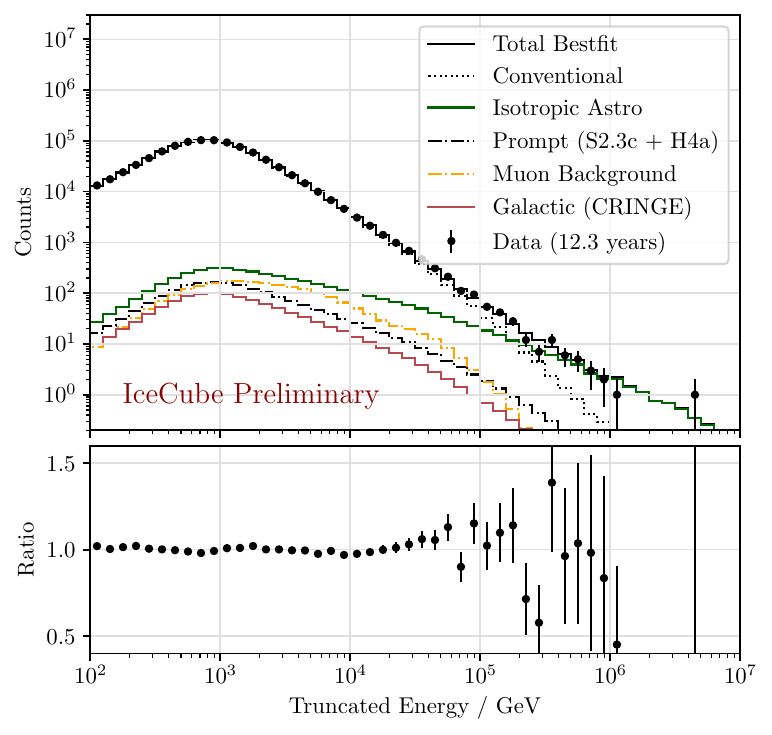}
    \caption{Data and bestfit results as function of the muon energy proxy. The dominating conventional atmospheric flux (black dotted line) is very similar to the total sum except for the highest energies, where the isotropic astrophysical signal (dark green) starts to dominate. The best-fit prefers no prompt atmospheric flux contribution, shown here is the prediction modeled with \textit{MCEq}, see Section \ref{sec:analysis} (black dashdotted). The muon background (orange) arises from mis-modeled atmospheric events which were misreconstructed as up-going. The best-fit galactic flux (dark red) contributes about 1.1\% to the total observed rate.}
    \label{fig:reco_energy}
\end{figure}

If the galactic component is neglected, we find a spectral index of $\gamma=2.40$ and the normalization increases by 6.5\%, see Fig. \ref{fig:contours}. The energy-projection of the analysis histogram with best-fit spectra is shown in Fig. \ref{fig:reco_energy}.

\subsection{Model differentiation}
As alternatives to our baseline model hypothesis \cite{cringe}, other calculations for galactic diffuse neutrino emission are tested: the Fermi-$\pi^0$ \cite{fermipi0} and KRA$_{\gamma}$ models \cite{KRAgamma}, which are based on Fermi-LAT data. The Fermi-$\pi^0$-model extrapolates the observed spectral index of $2.7$ as a single powerlaw. The KRA$_{\gamma}$ formulation assumes a radial dependence for cosmic ray diffusion, leading to a spectral hardening towards the galactic center. It contains a variable cutoff at 5(50) PeV in the galactic cosmic ray spectrum, yielding the KRA$_{\gamma}^{5(50)}$ models. Finally, we test for two analytic models following Fang\&Murase \cite{FangMuraseModels}, which assume a factorization of gas and cosmic ray density from the line of sight integral. The galactic disk is then homogeneously filled assuming the radial distribution to be either constant or to follow the distribution of supernova-remnants. For the spectrum we chose a single powerlaw analogously to Fermi-$\pi^0$ with a spectral index of $\gamma=2.7$. These models (FM-const and FM-SNR) result in predictions following the geometry of the galactic disk and are  independent of any gas structure.

\begin{figure}
    \centering
    \includegraphics[width=0.8\textwidth]{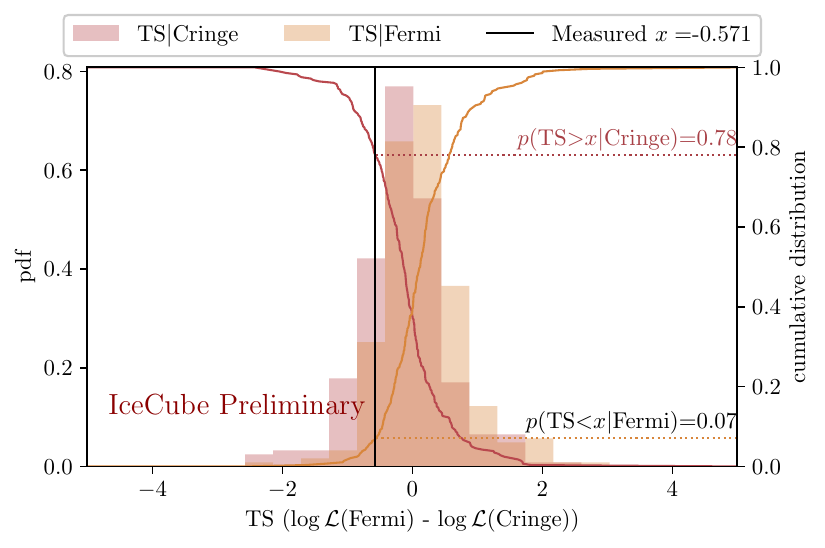}
    \caption{Example for a model differentiation test with the Fermi-$\pi^0$ model. Each $TS$ distribution is calculated from 600 pseudoexperiments. The quantiles are integrated from the left (right) for the alternative (baseline) model.}
    \label{fig:model_differentiation}
\end{figure}

We test all these alternative hypotheses against our baseline model by measuring the test-statistic $x = \log\mathcal{L}(H_{\mathrm{alternative}}) - \log\mathcal{L}(H_{\mathrm{Cringe}})$ for all models. To determine the significance, the $TS$-distribution is obtained from pseudo-experiments, which are drawn assuming one of the two hypotheses to be true, resulting in two distributions: $TS \vert \mathrm{Cringe} = \log\mathcal{L}(H_{\mathrm{alternative}} \vert \mathrm{Cringe}) - \log\mathcal{L}(H_{\mathrm{Cringe}}  \vert \mathrm{Cringe})$ and $TS \vert \mathrm{alternative}$. An example of this procedure is shown in Fig. \ref{fig:model_differentiation}. All resulting $p$-values from the tests are shown in Tab. \ref{tab:modeldifferentiation}. While we measure $x<0$ for all tests, it is currently not possible to establish a preferred model.

\begin{table}[]
    \centering
    \begin{tabular}{|c|c|c|c|c|}
    \hline
        Model & fitted $\Psi_{\mathrm{model}}$ & $x$ & $p(TS<x\vert \mathrm{Model})$ & $p(TS>x\vert \mathrm{Cringe})$ \\ \hline
        Fermi-$\pi^0$ & $4.7\pm^{+2}_{-2}$        & -0.571 & 0.07  & 0.78   \\ \hline
        KRA$_{\gamma}^{50}$ & $0.7\pm^{+0.4}_{-0.4}$    & -1.12  & 0.032 & 0.545  \\ \hline
        KRA$_{\gamma}^{5}$  & $1 \pm^{+0.5}_{-0.5}$     & -0.85  & 0.054 &  0.63  \\ \hline
        FM-SNR &  $1.6 \pm^{+0.8}_{-0.8}$ & -2.695 & 0.023 & 0.854  \\ \hline
        FM-const & $0.8\pm^{+0.6}_{-0.6}$ & -1.304 & 0.002 &  0.896 \\ \hline
    \end{tabular}
    \caption{Model differentiation tests. All models fit a nonzero galactic flux and measure $x<0$ (baseline hypothesis preferred). The significance of this preference ($p$-value) is calculated here only conditionally, assuming either the baseline or the alternative model to be true.}
    \label{tab:modeldifferentiation}
\end{table}

\begin{figure}
    \centering
    \includegraphics[width=1.\columnwidth]{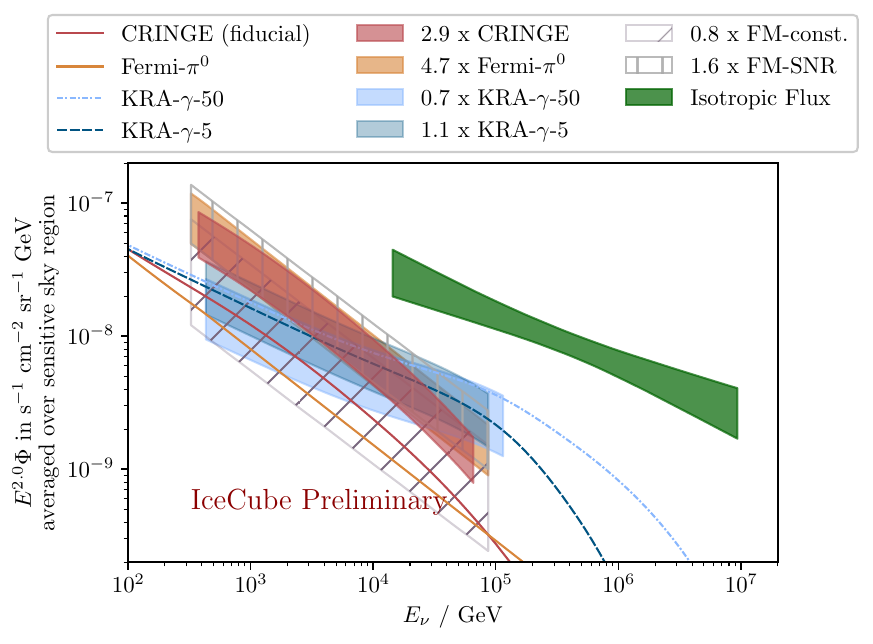}
    \caption{Unfolded Spectra of the two fitted neutrino flux contributions. The shown 68\% error bands are constructed by variation of the signal parameters inside the likelihood space. Model predictions for galactic diffuse emission which are based on gas maps are shown as lines.}
    \label{fig:spectra}
\end{figure}

\section{Conclusion}\label{sec:conclusion}
We present an updated measurement of the diffuse astrophysical neutrino flux using 12.3 years of IceCube track data, measuring an isotropic component described as a single powerlaw with the parameters $\phi_{\mathrm{isotropic}} = 1.51 \pm^{0.22}_{0.23}$ and $\gamma = 2.38\pm^{0.08}_{0.08}$. We observe a preference for a galactic contribution in the Northern Sky at the $2.7\sigma$ level. Removing this component increases the measured isotropical flux normalization by about 6.5\%. This result is consistent with the first observation of neutrino emission from the galactic plane, which is based on a dedicated selection of IceCube cascade events, and which fitted a very similar overall model normalization for the Fermi-$\pi^0$ model with a significance of $4.7\sigma$ \citep{dnncascades, mircosteve:2023icrc}. It is also consistent with a test for a galactic contribution using starting track events \cite{estes:2023icrc}. Although the difference is not statistically significant, it is interesting to note that the fitted model scales for the KRA-models are a factor $\approx 2$ larger in this Northern Sky measurement than in the all-sky cascade analysis.

For the Northern Sky, these measurements paint a consistent first picture of a neutrino flux from the galactic plane which is moderately stronger than the most recent model predictions for a diffuse-emission only scenario suggest (not considering contributions from unresolved sources) \cite{cringe}. However, obtaining any information on the spectral and spatial structure of this signal will require significantly more data, which could for example be achieved following ideas as outlined in \cite{pierpaolo:2023icrc} and \cite{globalfit:2023icrc} by moving towards global analyses of IceCube data.

\bibliographystyle{ICRC}
\bibliography{references}
\clearpage
\section*{Full Author List: IceCube Collaboration}

\scriptsize
\noindent
R. Abbasi$^{17}$,
M. Ackermann$^{63}$,
J. Adams$^{18}$,
S. K. Agarwalla$^{40,\: 64}$,
J. A. Aguilar$^{12}$,
M. Ahlers$^{22}$,
J.M. Alameddine$^{23}$,
N. M. Amin$^{44}$,
K. Andeen$^{42}$,
G. Anton$^{26}$
% ...
\\
\\
% ...
$^{12}$ Universit{\'e} Libre de Bruxelles, Science Faculty CP230, B-1050 Brussels, Belgium \\
% ...
$^{17}$ Department of Physics, Loyola University Chicago, Chicago, IL 60660, USA \\
$^{18}$ Dept. of Physics and Astronomy, University of Canterbury, Private Bag 4800, Christchurch, New Zealand \\
% ...
$^{22}$ Niels Bohr Institute, University of Copenhagen, DK-2100 Copenhagen, Denmark \\
$^{23}$ Dept. of Physics, TU Dortmund University, D-44221 Dortmund, Germany \\
% ...
$^{26}$ Erlangen Centre for Astroparticle Physics, Friedrich-Alexander-Universit{\"a}t Erlangen-N{\"u}rnberg, D-91058 Erlangen, Germany \\
% ...
$^{40}$ Dept. of Physics and Wisconsin IceCube Particle Astrophysics Center, University of Wisconsin{\textendash}Madison, Madison, WI 53706, USA \\
% ...
$^{42}$ Department of Physics, Marquette University, Milwaukee, WI, 53201, USA \\
% ...
$^{44}$ Bartol Research Institute and Dept. of Physics and Astronomy, University of Delaware, Newark, DE 19716, USA \\
% ...
$^{63}$ Deutsches Elektronen-Synchrotron DESY, Platanenallee 6, 15738 Zeuthen, Germany  \\
$^{64}$ Institute of Physics, Sachivalaya Marg, Sainik School Post, Bhubaneswar 751005, India \\
% ...

\subsection*{Acknowledgements}

\noindent
The authors gratefully acknowledge the support from the following agencies and institutions:
USA {\textendash} U.S. National Science Foundation-Office of Polar Programs,
U.S. National Science Foundation-Physics Division,
U.S. National Science Foundation-EPSCoR,
Wisconsin Alumni Research Foundation,
Center for High Throughput Computing (CHTC) at the University of Wisconsin{\textendash}Madison,
Open Science Grid (OSG),
Advanced Cyberinfrastructure Coordination Ecosystem: Services {\&} Support (ACCESS),
Frontera computing project at the Texas Advanced Computing Center,
U.S. Department of Energy-National Energy Research Scientific Computing Center,
Particle astrophysics research computing center at the University of Maryland,
Institute for Cyber-Enabled Research at Michigan State University,
and Astroparticle physics computational facility at Marquette University;
Belgium {\textendash} Funds for Scientific Research (FRS-FNRS and FWO),
FWO Odysseus and Big Science programmes,
and Belgian Federal Science Policy Office (Belspo);
Germany {\textendash} Bundesministerium f{\"u}r Bildung und Forschung (BMBF),
Deutsche Forschungsgemeinschaft (DFG),
Helmholtz Alliance for Astroparticle Physics (HAP),
Initiative and Networking Fund of the Helmholtz Association,
Deutsches Elektronen Synchrotron (DESY),
and High Performance Computing cluster of the RWTH Aachen;
Sweden {\textendash} Swedish Research Council,
Swedish Polar Research Secretariat,
Swedish National Infrastructure for Computing (SNIC),
and Knut and Alice Wallenberg Foundation;
European Union {\textendash} EGI Advanced Computing for research;
Australia {\textendash} Australian Research Council;
Canada {\textendash} Natural Sciences and Engineering Research Council of Canada,
Calcul Qu{\'e}bec, Compute Ontario, Canada Foundation for Innovation, WestGrid, and Compute Canada;
Denmark {\textendash} Villum Fonden, Carlsberg Foundation, and European Commission;
New Zealand {\textendash} Marsden Fund;
Japan {\textendash} Japan Society for Promotion of Science (JSPS)
and Institute for Global Prominent Research (IGPR) of Chiba University;
Korea {\textendash} National Research Foundation of Korea (NRF);
Switzerland {\textendash} Swiss National Science Foundation (SNSF);
United Kingdom {\textendash} Department of Physics, University of Oxford.
\end{document}